# A Time-independent Way to Probe $D^0 - \bar{D}^0$ Mixing at $\tau$-charm Factories

Zhi-zhong Xing [1]

*Sektion Physik, Theoretische Physik, Universität München,*

*Theresienstrasse 37, D-80333 München, Germany*

*and*

*Department of Physics, Faculty of Science, Nagoya University,*

*Chikusa-ku, Nagoya 464-01, Japan* [2]

## Abstract

$D^0 - \bar{D}^0$ mixing leads to the mass and width differences in the mass eigenstates of $D^0$ and $\bar{D}^0$ mesons (measured by parameters $x_D$ and $y_D$ respectively), but their magnitudes cannot be reliably predicted by the standard model. We show that it is possible to separately determine $x_D$ and $y_D$ through *time-integrated* measurements of the dilepton events of coherent $D^0 \bar{D}^0$ decays on the $\psi(4.14)$ resonance at a $\tau$-charm factory.

---

[1]Electronic address: xing@eken.phys.nagoya-u.ac.jp

[2]Mailing address

It is known in charm physics that mixing between $D^0$ and $\bar{D}^0$ mesons can arise naturally, since both of them couple to a subset of the virtual and real intermediate states. The rate of $D^0 - \bar{D}^0$ mixing is commonly measured by two well-defined parameters $x_D$ and $y_D$, which correspond to the mass and width differences in the mass eigenstates of $D^0$ and $\bar{D}^0$. In the standard model the short-distance (box diagram) contribution to $D^0 - \bar{D}^0$ mixing is expected to be negligibly small [1], but different approaches to the long-distance effects have given dramatically different estimates for the magnitudes of $x_D$ and $y_D$ [2, 3, 4, 5].

The latest E691 data of Fermilab fixed target experiments give an upper bound on $D^0 - \bar{D}^0$ mixing: $r_D \approx (x_D^2 + y_D^2)/2 < 0.37\%$ [6]. If $x_D$ and $y_D$ are well below $10^{-2}$ as predicted by the dispersive approach of ref. [2] or the heavy quark effective theory of ref. [4], then the observation of $r_D$ at the level of $10^{-4}$ or so will imply the presence of new physics [7, 8, 9, 10].

Today much more theoretical effort is needed to make sure of the order of $x_D$ and $y_D$. On the experimental side, searches for $D^0 - \bar{D}^0$ mixing to a high degree of accuracy (e.g., $r_D \sim 10^{-4}$ to $10^{-5}$) are expected to be available at future fixed target facilities, $B$-meson factories and $\tau$-charm factories [11, 12, 13].

In this work we investigate how the subtlety of $D^0 - \bar{D}^0$ mixing can be probed at a $\tau$-charm factory. We show that it is possible to separately determine the mixing parameters $x_D$ and $y_D$ by use of the *time-integrated* dilepton events of coherent $D^0 \bar{D}^0$ decays at the $\psi(4.14)$ resonance. The importance of such measurements is that knowledge of the relative magnitude of $x_D$ and $y_D$ can definitely clarify the ambiguities of current theoretical estimates and shed light on possible sources of new physics in $D^0 - \bar{D}^0$ mixing.

In the presence of $CP$ violation, the $CP$ eigenstates $|D_1\rangle \equiv (|D^0\rangle + |\bar{D}^0\rangle)/\sqrt{2}$ and $|D_2\rangle \equiv (|D^0\rangle - |\bar{D}^0\rangle)/\sqrt{2}$ are related to the mass eigenstates $|D_{\rm L}\rangle$ and $|D_{\rm H}\rangle$ by

$$
\begin{aligned}
|D_{\rm L}\rangle &= \cosh(\mathrm{i}\phi)|D_1\rangle - \sinh(\mathrm{i}\phi)|D_2\rangle , \\
|D_{\rm H}\rangle &= \cosh(\mathrm{i}\phi)|D_2\rangle - \sinh(\mathrm{i}\phi)|D_1\rangle ,
\end{aligned} \quad (1)
$$

where the subscripts "L" and "H" stand for Light and Heavy respectively, and $\phi$ is a complex phase. The proper-time evolution of an initially ($t=0$) pure $D^0$ or $\bar{D}^0$ is given as

$$
\begin{aligned}
|D^0(t)\rangle &= f_0(t)\left[f_+(t)|D^0\rangle + \exp(+\mathrm{i}2\phi)f_-(t)|\bar{D}^0\rangle\right] , \\
|\bar{D}^0(t)\rangle &= f_0(t)\left[f_+(t)|\bar{D}^0\rangle + \exp(-\mathrm{i}2\phi)f_-(t)|D^0\rangle\right] ,
\end{aligned} \quad (2)
$$

in which the evolution functions read $f_0(t) = \exp[-(\mathrm{i}m + \Gamma/2)t]$, $f_+(t) = \cosh[(\mathrm{i}\Delta m - \Delta\Gamma/2)t/2]$ and $f_-(t) = \sinh[(\mathrm{i}\Delta m - \Delta\Gamma/2)t/2]$. Here we have used the common notations $m \equiv (m_\mathrm{L} + m_\mathrm{H})/2$, $\Gamma \equiv (\Gamma_\mathrm{L} + \Gamma_\mathrm{H})/2$, $\Delta m \equiv m_\mathrm{H} - m_\mathrm{L}$ and $\Delta\Gamma \equiv \Gamma_\mathrm{L} - \Gamma_\mathrm{H}$. Furthermore, we define $x_D \equiv \Delta m/\Gamma$ and $y_D \equiv \Delta\Gamma/(2\Gamma)$ as two characteristic parameters of $D^0 - \bar{D}^0$ mixing.

For fixed target experiments or $e^+e^-$ collisions at the $\Upsilon(4S)$ resonance, the $D^0$ and $\bar{D}^0$ mesons can be produced incoherently. Knowledge of $D^0 - \bar{D}^0$ mixing is expected to come from ratios of the wrong-sign to right-sign events of semileptonic $D$ decays:

$$r \equiv \frac{\Gamma(D^0 \to l^- X^+)}{\Gamma(D^0 \to l^+ X^-)} = \exp(-4\mathrm{Im}\phi) \frac{1-\alpha}{1+\alpha} , \tag{3a}$$

$$\bar{r} \equiv \frac{\Gamma(\bar{D}^0 \to l^+ X^-)}{\Gamma(\bar{D}^0 \to l^- X^+)} = \exp(+4\mathrm{Im}\phi) \frac{1-\alpha}{1+\alpha} , \tag{3b}$$

where $\alpha = (1 - y_D^2)/(1 + x_D^2)$. Note that nonvanishing $\mathrm{Im}\phi$ signifies $CP$ violation in $D^0 - \bar{D}^0$ mixing. To fit more accurate data in the near future, we prefer the following mixing parameter:

$$r_D \equiv \frac{r + \bar{r}}{2} = \cosh(4\mathrm{Im}\phi) \frac{1-\alpha}{1+\alpha} . \tag{4}$$

For $\mathrm{Im}\phi \sim 1\%$, the value of $\cosh(4\mathrm{Im}\phi)$ deviates less than 0.1% from unity. Thus this overall factor of $r_D$ is safely negligible [3]. The latest E691 data [6] give $r \approx \bar{r} \approx r_D \approx (x_D^2 + y_D^2)/2 < 0.37\%$ for small $x_D$ and $y_D$, where $\exp(-4\mathrm{Im}\phi) \approx \exp(+4\mathrm{Im}\phi) \approx 1$, a worse approximation than $\cosh(4\mathrm{Im}\phi) \approx 1$, has been used.

For a $\tau$-charm factory running at the $\psi(4.14)$ resonance, the coherent $D^0\bar{D}^0$ events can be produced through $\psi(4.14) \to \gamma (D^0\bar{D}^0)_{C=+}$ or $\psi(4.14) \to \pi^0 (D^0\bar{D}^0)_{C=-}$, where $C$ stands for the charge-conjugation parity [7, 13]. The time-dependent wave function for a $(D^0\bar{D}^0)_C$ pair at rest is written as

$$\frac{1}{\sqrt{2}} \left[ |D^0(\mathbf{k}, t)\rangle \otimes |\bar{D}^0(-\mathbf{k}, t)\rangle + C|D^0(-\mathbf{k}, t)\rangle \otimes |\bar{D}^0(\mathbf{k}, t)\rangle \right] , \tag{5}$$

where $\mathbf{k}$ is the three-momentum vector of $D^0$ and $\bar{D}^0$ mesons. For our purpose, we only consider the primary dilepton events which are directly emitted from the coherent $(D^0\bar{D}^0)_C$ decays. Let $N_C^{\pm\pm}$ and $N_C^{+-}$ denote the time-integrated numbers of like-sign and opposite-sign dilepton events, respectively. After a straightforward calculation [7, 14, 15], we obtain

$$N_C^{++} = N_C \exp(+4\mathrm{Im}\phi) \left[ \frac{1 + Cy_D^2}{(1-y_D^2)^2} - \frac{1 - Cx_D^2}{(1+x_D^2)^2} \right] , \tag{6a}$$

---

[3] Indeed the magnitude of $\mathrm{Im}\phi$ can be determined by measurements of the $CP$ asymmetry $(\bar{r} - r)/(\bar{r} + r)$, which is equal to $\sinh(4\mathrm{Im}\phi)/\cosh(4\mathrm{Im}\phi) \approx 4\mathrm{Im}\phi$.

$$N_C^{--} = N_C \exp(-4\mathrm{Im}\phi) \left[ \frac{1 + Cy_D^2}{(1 - y_D^2)^2} - \frac{1 - Cx_D^2}{(1 + x_D^2)^2} \right] , \qquad (6b)$$

and

$$N_C^{+-} = 2N_C \left[ \frac{1 + Cy_D^2}{(1 - y_D^2)^2} + \frac{1 - Cx_D^2}{(1 + x_D^2)^2} \right] , \qquad (7)$$

where $N_C$ is the normalization factor proportional to the rates of semileptonic $D^0$ and $\bar{D}^0$ decays. It is easy to check that the relation

$$N_-^{++} N_+^{--} = N_+^{++} N_-^{--} \qquad (8)$$

holds stringently, and it is independent of the magnitudes of $D^0 - \bar{D}^0$ mixing and $CP$ violation.

Note that a coherent $D^0 \bar{D}^0$ pair with $C = -$ can be straightforwardly produced from the decay of the $\psi(3.77)$ resonance [10, 13]. Its time-independent decay rates to the like-sign and opposite-sign dileptons obey eqs. (6) and (7), respectively. At a $\tau$-charm factory the $(D^0 \bar{D}^0)_{C=-}$ events at both the $\psi(3.77)$ and $\psi(4.14)$ resonances will be accumulated, and a combination of them may increase the sensitiveness of our approach to probing $D^0 - \bar{D}^0$ mixing.

Usually one is interested in the following two types of observables:

$$a_C \equiv \frac{N_C^{++} - N_C^{--}}{N_C^{++} + N_C^{--}} , \qquad r_C \equiv \frac{N_C^{++} + N_C^{--}}{N_C^{+-}} , \qquad (9)$$

which signify nonvanishing $CP$ violation and $D^0 - \bar{D}^0$ mixing, respectively. Explicitly, we find

$$a_- = a_+ = \frac{\sinh(4\mathrm{Im}\phi)}{\cosh(4\mathrm{Im}\phi)} \approx 4\mathrm{Im}\phi \qquad (10)$$

for small $\mathrm{Im}\phi$. If $\mathrm{Im}\phi$ is at the level of $10^{-3}$ or larger, it can be measured to 3 standard deviations at the second-round experiments of a $\tau$-charm factory with about $10^7$ like-sign dileptons (or equivalently, about $10^{10}$ $D^0 \bar{D}^0$ events). Furthermore,

$$r_- = \cosh(4\mathrm{Im}\phi) \frac{1 - \alpha}{1 + \alpha} , \qquad r_+ = \cosh(4\mathrm{Im}\phi) \frac{\beta - \alpha^2}{\beta + \alpha^2} , \qquad (11)$$

where $\beta = (1 + y_D^2)/(1 - x_D^2)$. One can see that $r_- = r_D$ holds without any approximation. For small $x_D$ and $y_D$, we have $r_- \approx (x_D^2 + y_D^2)/2$ and $r_+ \approx 3r_-$. These two approximate results are well-known in the literature (see, e.g., [7]). In such an approximation, however, the relative size of $x_D^2$ and $y_D^2$ cannot be determined.

To distinguish between the different contributions of $x_D$ and $y_D$ to $D^0 - \bar{D}^0$ mixing, one has to measure $r_\pm$ as precisely as possible. With the help of eq. (11), we show that the magnitudes of $x_D$ and $y_D$ can be separately determined as follows:

$$x_D^2 = \left(\frac{1+r_-}{1-r_-}\cdot\frac{1+3r_-}{1-r_-} - \frac{1+r_+}{1-r_+}\right)\left(\frac{1+r_-}{1-r_-} - \frac{1+r_+}{1-r_+}\right)^{-1}, \quad (12a)$$

$$y_D^2 = \left(\frac{1-r_-}{1+r_-}\cdot\frac{1-3r_-}{1+r_-} - \frac{1-r_+}{1+r_+}\right)\left(\frac{1-r_+}{1+r_+} - \frac{1-r_-}{1+r_-}\right)^{-1}. \quad (12b)$$

Here it is worth emphasizing that $\cosh(4\text{Im}\phi)$ as the overall (and common) factor of $r_D$, $r_-$ and $r_+$ can be safely neglected. In the approximations up to $O(r_-^2)$ and $O(r_+^2)$, we obtain two simpler relations:

$$x_D^2 - y_D^2 \approx 2\,\frac{r_+ - 3r_-}{r_+ - r_-}, \qquad x_D^2 + y_D^2 \approx 4r_-\,\frac{r_+ - 2r_-}{r_+ - r_-}. \quad (13)$$

Thus it is crucial to examine the deviation of the ratio $r_+/r_-$ from 3, in order to find the difference between $x_D^2$ and $y_D^2$. Instructively, we consider three special cases:

$$x_D \gg y_D \quad \Longrightarrow \quad \frac{r_+}{r_-} \approx 3 + 2r_- > 3, \quad (14a)$$

$$x_D \approx y_D \quad \Longrightarrow \quad \frac{r_+}{r_-} \approx 3 - 9r_-^2 \approx 3, \quad (14b)$$

$$x_D \ll y_D \quad \Longrightarrow \quad \frac{r_+}{r_-} \approx 3 - 2r_- < 3. \quad (14c)$$

These relations can be directly derived from eq. (11) or (12). If $r_-$ is close to the current experimental bound (i.e., $r_- = r_D \approx (x_D^2 + y_D^2)/2 < 0.37\%$), then measurements of $r_+/r_-$ to the accuracy of $10^{-4}$ can definitely establish the relative magnitude of $x_D$ and $y_D$. To this goal, about $10^8$ like-sign dileptons (or equivalently, about $10^{11}$ events of $(D^0\bar{D}^0)_{C=-}$ and $(D^0\bar{D}^0)_{C=+}$ pairs) are needed. Certainly such a measurement can only be carried out at the second-generation $\tau$-charm factories (beyond the ones under consideration at present).

In principle, there is another possibility to determine $x_D^2$ and $y_D^2$ separately. If the production cross-sections of $\psi(4.14) \to \gamma(D^0\bar{D}^0)_{C=+}$ and $\psi(4.14) \to \pi^0(D^0\bar{D}^0)_{C=-}$ can be reliably predicted or measured, it is possible to fix the ratio of the two normalization factors (i.e., $n \equiv N_-/N_+$) independent of the dilepton events shown in eqs. (6) and (7). Then an asymmetry between the $C=-$ and $C=+$ opposite-sign dilepton events can be obtained as follows:

$$\Delta^{+-} \equiv \frac{N_+^{+-} - N_-^{+-}}{N_+^{+-} + N_-^{+-}} \approx \frac{1-n}{1+n} - \frac{3-n}{1+n}\cdot\frac{x_D^2 - y_D^2}{2}. \quad (15)$$

We see that $n=1$ is the most favorable case (i.e., $(D^0\bar{D}^0)_{C=+}$ and $(D^0\bar{D}^0)_{C=-}$ events have the same production rate at the $\psi(4.14)$ resonance), in which $\Delta^{+-}$ directly measures the difference between $x_D^2$ and $y_D^2$. A comparison between $\Delta^{+-}$ and $r_-$ (or $r_+$) is able to determine $x_D^2$ and $y_D^2$ separately. In this way $r_-$ need not be measured as precisely as in the first approach discussed above, however, the accurate value of $n$ is necessary. Unfortunately, it seems impossible at present to precisely determine $n$ from either theory or experiments.

We have shown that it is possible to separately determine the $D^0 - \bar{D}^0$ mixing parameters $x_D$ and $y_D$ by time-integrated measurements of the dilepton events of $(D^0\bar{D}^0)_{C=\pm}$ decays at a $\tau$-charm factory. In the assumption of a dedicated accelerator running for one year at an average luminosity of $10^{33}\text{s}^{-1}\text{cm}^{-2}$, about $10^7$ events of $\gamma(D^0\bar{D}^0)_{C=+}$ and the similar number of $\pi^0(D^0\bar{D}^0)_{C=-}$ are expected to be produced at the $\psi(4.14)$ resonance [13] [4]. The precision of $10^{-4}$ to $10^{-5}$ in measurements of $r_-$ and $r_+$ is achievable if one assumes zero background and enough running time [13, 16]. To measure the ratio $r_+/r_-$ up to the accuracy of $10^{-4}$, however, much experimental effort is needed. If $D^0 - \bar{D}^0$ mixing were at the level of $r_D \sim 10^{-3}$ (or at least $r_D \geq 10^{-4}$), then the relative magnitude of $x_D$ and $y_D$ should be detectable in the second-round experiments of a $\tau$-charm factory.

I would like to thank H. Fritzsch and A.I. Sanda for their warm hospitality during my research stay in München and Nagoya, respectively. I am also grateful to D. M. Kaplan and T. Liu for some useful discussions. This work was supported by both the Alexander von Humboldt Foundation and the Japan Society for the Promotion of Science.

---

[4] A rough estimate gives $n \approx 0.78$. Note that more $(D^0\bar{D}^0)_{C=-}$ events can be produced at the $\psi(3.77)$ resonance, but they are only applicable to the measurement of $r_-$.